\documentclass{ws-ijmpcs}

\begin{document}

\markboth{JinLin Han et al. }
{Sino-German $\lambda$6cm polarization survey of the Galactic plane: a summary}

%
\catchline{}{}{}{}{}
%

\title{THE SINO-GERMAN $\lambda$6\ CM POLARIZATION SURVEY OF THE GALACTIC PLANE: A SUMMARY 
}

\author{J. L.~HAN$^{1}$, W.~REICH$^{2}$, X. H.~SUN$^{1,2}$, X. Y.~GAO$^{1}$, L.~XIAO$^{1}$, W. B. Shi$^{1}$,\\
 P.~REICH$^{2}$, R.~WIELEBINSKI$^{2}$}
\address{1. National Astronomical Observatories, Chinese Academy of Sciences,\\ 
Jia-20 Datun Road, Chaoyang District, Beijing 100012, China\\ 
hjl@nao.cas.cn \\ 
2. Max-Planck-Institut f\"{u}r Radioastronomie, \\ Auf dem H\"{u}gel 69, 
53121 Bonn, Germany\\
wreich@mpifr-bonn.mpg.de
}

\maketitle


\begin{abstract}
We have finished the $\lambda$6\ cm polarization survey of the
Galactic plane using the Urumqi 25\ m radio telescope.  It covers
$10^{\circ} \leq l \leq 230^{\circ}$ in Galactic longitude and $|b|
\leq 5^{\circ}$ in Galactic latitude. The new polarization maps not
only reveal new properties of the diffuse magnetized interstellar
medium, but also are very useful for studying individual objects such
as {\sc Hii} regions, which may act as Faraday screens with strong
regular magnetic fields inside, and supernova remnants for their 
polarization properties and spectra. The high sensitivity of the
survey enables us to discover two new SNRs G178.2$-$4.2 and
G25.3$-$2.1 and a number of {\sc Hii} regions.

\end{abstract}

\ccode{PACS numbers: 98.38.-j, 98.38.Mz, 98.58.Hf}

\section{Introduction}

A polarization survey of diffuse Galactic emission provides a direct
image of the transverse magnetic field, if Faraday rotation is
negligible. The Sino-German $\lambda$6\ cm polarization survey of the
Galactic plane was set as the main task for the partner group of the
Max-Planck-Institut f\"{u}r Radioastronomie (MPIfR) at the National
Astronomical Observatories of Chinese Academy of Sciences (NAOC),
which was commissioned in 2001.

Previously the Effelsberg 100\ m telescope was used to map the
Galactic polarized emission up to medium latitudes\cite{ufr+99,rfr+04}
at 1.4~GHz with an angular resolution of $9.4'$. Part of the data were
used to compensate the missing large-scale structures in the Canadian
Galactic Plane Survey (CGPS)\cite{lrr+10} observed with the DRAO
synthesis telescope at $1'$ angular resolution. At 1.4~GHz, mainly
polarized emission from the local interstellar medium is observed
\cite{gdm+01,ufr+99}. The polarization survey by the Westerbork
Synthesis Radio Telescope at 350~MHz\cite{hkd03a,hkd03b} penetrates
through an even shorter distance into the magneto-ionic
medium. Earlier 2.7~GHz Galactic plane polarization surveys were made
with the Effelsberg 100\ m telescope\cite{jfr87,drrf99} at $4.3'$
resolution tracing more distant emission. The Galactic plane in the
southern sky was surveyed by the Parkes telescope at
2.4~GHz\cite{dhjs97}, while sections of the Galactic plane were
observed with the Australian Telescope Compact Array at 1.4~GHz with a
high angular resolution\cite{gdm+01,hgm+06}. The observing
frequencies of previous polarization surveys are all lower than 3~GHz,
which implies a small distance to the polarization horizon. In other
words, the results suffer significant depolarization for distant
emission. To penetrate even deeper into the magneto-ionic medium,
observations at higher frequencies are required.

Using the Urumqi 25\ m radio telescope, the Sino-German $\lambda$6~cm
polarization survey covers the Galactic plane from $10^{\circ}\le
l\le230^{\circ}$ and $|b|\le5^{\circ}$ with an angular resolution of
$9.5'$, which enables us to see through the complex magneto-ionized
structures even for a $RM$ of several hundred rad\ m$^{-2}$. In the
inner Galaxy of $10^{\circ} \le l \le 60^{\circ}$, the radio emission
at this frequency comes from diffuse emission and individual objects
of a few spiral arms\cite{hhs09}. Towards the direction of $l \sim
80^{\circ}$, the line of sight is tangential to the local arm, and the
large-scale magnetic fields run parallel to the line of sight. There
are many objects and structures of different distances in this
projected area, forming a radio complex. In the Galactic anti-center
region ($120^{\circ} \le l \le 230^{\circ}$), the emission comes from
the Perseus arm and maybe also a weaker outer arm, where the
large-scale magnetic fields run almost perpendicular to the line of
sight.

\section{Instrumental setup and commissioning}

The Sino-German $\lambda$6\ cm polarization survey of the Galactic
plane has been conducted using the Urumqi 25\ m telescope of Xinjiang
(former Urumqi) Astronomical Observatory, Chinese Academy of
Sciences. The telescope is located about 70\ km south of Urumqi city
at an altitude of 2029\ m above sea level with the geographic
longitude 87$^{\circ}$E and latitude 43$^{\circ}$N.

The dual-channel $\lambda$6\ cm receiver was designed by Mr. Otmar
Lochner and constructed at MPIfR in Germany, based on a receiver used
at the Effelsberg 100 m telescope. The new receiver has a higher
stability and a lower 1/f-noise. In 2004, the receiver was installed
at the secondary focus of the Urumqi 25\ m telescope. The $\lambda$6
cm system includes a corrugated feed horn and an orthogonal transducer
to convert the signal into left-hand (L) and right-hand (R)
polarization channels.  The signal of the two channels is amplified by
cooled HEMT pre-amplifiers working below 15\ K in a dewar and then
mixed with the local oscillator signal. Two splitters follow, which
split one set of L and R channels into two sets, one used to directly
detect the power of $L$=LL* and $R$=RR* for Stokes $I$, and the other
one is feed into the IF polarimeter for producing LR* and RL* for the
Stokes $U$ and $Q$. The voltage-frequency converters in the focus
cabin are used for converting the detected voltage signals to
frequencies, which enables the long-distance transportation of the
signals to a digital backend in the observatories control room.  A
MPIfR pocket backend counts the frequency-coded signals from the four
polarization channels (RR*, LL*, RL*, LR*) at a sampling rate of
32\ msec. Every 32\ msec the frontend setting changes, so that either
a calibration signal of 1.7\ K $\rm T_{a}$ is added to the antenna
signal and/or the signal phase is shifted by $180^{\circ}$.  For two
subsequent phases within a cycle, the calibration signal is switched
on, so that any gain changes of the system can be monitored. The phase
of the output signals is switched by $180^{\circ}$ alternatively for
every phase, so that the quadratic terms of the polarimeter are
canceled. In one whole cycle of 128 ms, four combinations of different
settings of calibration and phase-switching are realized.

\begin{table}
\centering
Table 1. Survey parameters of the Urumqi $\lambda$6\ cm system\\
\begin{tabular}{ll}\hline
  Frequency [MHz] & 4800/4963 \\
  Bandwidth [MHz] & 600/295 \\
  T$_{\rm sys}$ [K] T$_a$& 22 \\
  HPBW [$'$] & 9.5 \\
  Scan direction & $l$ and $b$ \\
                 & RA and DEC \\
  T$_{B}$/S (Gain)                            & 0.164 K/Jy        \\
  First side lobe                             & $< 2\%$      \\
  Instrumental polarization                   & $< 2\%$      \\
  $\sigma_I$ [mK] T$_b$& 1.0 \\
  $\sigma_{PI}$ [mK]T$_b$ & 0.4 \\ \hline
  \end{tabular}
\end{table}

The raw data are transfered to a Linux PC via the TCP/IP protocol and
stored on disk for further processing in the MBFITS format developed
for the Effelsberg 100\ m telescope.  By using the TOOLBOX-software
package, the raw data of the four backend channels (RR*, LL*, RL*,
LR*) are converted from a time series into a tabulated format with a
user specified spatial separation on the sky using a
sinc-interpolation function. The calibration signals of each subscan
(row or column) of a map are extracted and fitted to account for any
gain drifting and also to control the phase stability of the IF
polarimeter. The tabulated maps are then transformed into NOD2
maps\cite{has74} with the polarization U and Q-channels corrected for
parallactic angles. The Linux PC is also used to command the telescope
control computer to move the telescope in a number of different
astronomical coordinate systems (RA/DEC or $l/b$) and in the
azimuth/elevation coordinate system of the telescope.

Test observations were made in August 2004, and the relevant system
and antenna parameters were determined (see Table 1). By periodically
injecting a known signal, we got the system temperature measured as
22~K\ T$_{a}$ when the $\lambda$6\ cm receiver is pointing to the
zenith of a clear sky. The strong radio source 3C~286 is used to
determine other parameters, such as the beam size, effective
collecting area, main beam efficiency and gain. Indian satellites
which are known to emit strong in the $\lambda$6\ cm band were
observed to estimate the instrumental effects, e.g. the side
lobes. The unpolarized radio source 3C~295 was used to calculate the
leakage of the total intensity into the polarization channels, namely
the instrumental polarization. From these tests, we found the
$\lambda$6\ cm beam is rather symmetric and the first side lobes, are
less than 2\% of the main beam. Cross-scans were made in two
orthogonal directions to the source 3C~286, and showed that the
pointing accuracy of the system is stable and the offset is negligible
in comparison with the beam size.

During the test observations, we identified internal and external
interferences. TThe observatory's pulsar filter-bank backend
introduced the major internal interference, and was switched off
during the $\lambda$6\ cm observations. The external interferences
came from the ground radiation and the Indian satellites. The Nanshan
mountains around the site induce significant ground radiation picked
up by the sidelobes of the telescope, which produce a non-flat
baseline for scans of low elevations\cite{whs+07}. For the required
survey quality, observations must be done at elevations larger than
20$^{\circ}$, so that the baseline from the ground radiation can be
well subtracted by a first-order or second-order polynomial fit. There
are four locations of Geostationary Indian satellites, which emit
strong signals in the $\lambda$6 cm band, were observed at the azimuths
and elevations of (222.61$^{\circ}$, 30.34$^{\circ}$),
(198.53$^{\circ}$, 38.19$^{\circ}$), (186.06$^{\circ}$,
39.7$^{\circ}$), and (170.85$^{\circ}$, 39.4$^{\circ}$), which
emit strong signals up to 4810~MHz with a flux density of
$>$60\,000~Jy.  Our $\lambda$6~cm system suffers from their signals
even tens of degrees away via telescope side lobes. In November 2005,
the receiver was equipped with a switched bandpass filter. When it is
switched on, the lower band is cut off, and the central observing
frequency is shifted from 4800~MHz to 4963~MHz, and the bandwidth is
cut from 600~MHz to 295~MHz. We have to use this narrow band to
observe the southern parts of the Galactic plane.

\section{Observations and data reduction}

The $\lambda$6\ cm system is very sensitive， recording, for instance,
any vibration of the thin RF-transparent membrane of of the
focus-cabin by wind or small temperature variations in the focus
cabin, which was fixed. However, effects by a bird or clouds
passing the telescope beam and, in particular, emission from the sun
picked up by the telescope side lobes could not be fixed.  Therefore,
the survey observations were always carried out in clear nights.

The $\lambda$6\ cm survey observations started in August 2004, and
finished in April 2009. A very large section of the Galactic plane of
2\,200 square degrees for $10^{\circ} \leq l \leq230^{\circ}$ and $|b|
\leqslant 5^{\circ}$ has been scanned in two orthogonal directions:
along the Galactic longitude for the $l$ map and along the Galactic
latitude for the $b$ map. Considering the stability of the system and
the variable ground radiation, scans must be done for a finite size in
a given range of the sky. For a $l$ map, the typical map size of $l
\times b$ is $8.6^{\circ} \times 2.6^{\circ}$; for a $b$ map it is
$2.2^{\circ} \times 10.0^{\circ}$. There is always a small overlap of
the different patches. Before and after observations for each map,
calibrators are observed. 3C~286 served as our main calibrator. It has
an integrated flux density of $S_{\rm 6cm} = 7.5$~Jy at
$\lambda$6\ cm, 11.3\% polarization fraction, and a polarization angle
of 33$^{\circ}$.  3C~48 and 3C~138 were used as secondary
calibrators. The unpolarized radio source 3C~295 was also frequently
observed for the cleaning of the instrumental polarization.

As mentioned above, the original survey data were stored in a Linux
PC.  We used the ``toolbox'' and ``mapax'' software originally
developed at the MPIfR for the data reduction of Effelsberg continuum
and polarization observations. There are several steps in data
processing. The maps were first edited to remove spiky interference
and obvious very bad scans due to interference. Then we corrected
baseline curvatures by polynomial fitting and suppressed scanning
effects by applying an ``unsharp masking method''\cite{sr79}.  All
individual maps were therefore on a relative baselevel, where the two
ends of each subscan were set to zero. The instrumental polarization,
mainly the leakage of total power emission into the polarization
channels, was determined by observations of the unpolarized
calibrators 3C~295 and 3C~147, and corrected using the ``REBEAM"
method\cite{sris87}. Then the intensity calibrations was made to each
map. Finally, we weaved the processed $l$ maps and $b$ maps together
by using the "PLAIT" method\cite{eg88}, which further reduced
remaining scanning effects, to form the Stokes $I$, $U$, and $Q$ survey
maps. The survey strategy together with the results from a test
region of $122^{\circ}\leq l\leq129^{\circ}$ was published in the
first survey paper\cite{shr+07}.

\begin{figure}[pb]
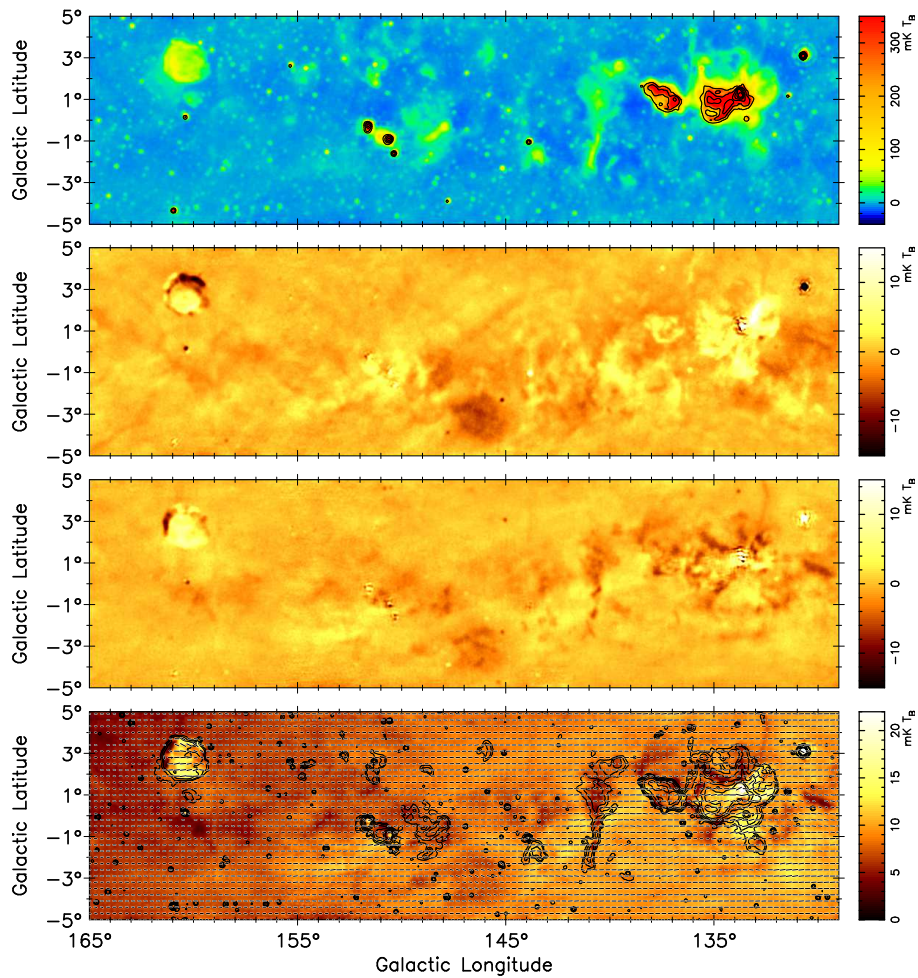

\centerline{\psfig{file=i.ps,bb=230 32 430 794,clip=true,angle=270,width=12cm}}\vspace{-1mm}
\centerline{\psfig{file=u.ps,bb=230 32 430 794,clip=true,angle=270,width=12cm}}\vspace{-1mm}
\centerline{\psfig{file=q.ps,bb=230 32 430 794,clip=true,angle=270,width=12cm}}\vspace{-1mm}
\centerline{\psfig{file=pi.ps,bb=230 32 467 794,clip=true,angle=270,width=12cm}}
\caption{A section of survey map. From top to bottom are the observed
  total intensity $I$ map, and Stokes $U$ and $Q$ maps.  The $PI$ map
  includes the large scale emission with $PA+90^{\circ}$ vectors and
  contours of total power overlaid.
\label{fig1}}
\end{figure}

\section{Results}

For every sky area, in general, we performed observations for
more than one $l$ map and $b$ map. Combining all observations
together, we obtained the $I$, $U$, and $Q$ maps for the whole survey
region.  The r.m.s.-noise in the $I$ map is about 1.0~mK T$_{b}$, and
in the $U$ and $Q$ maps it is about 0.5~mK T$_{b}$.  All survey data have
been released at the end of 2010\footnote{\sf
    http://zmtt.bao.ac.cn/6cm/}. See Fig.~1 for a section of survey
maps.  It is by far the highest frequency in a ground-based
polarization Galactic plane survey.

\subsection{Survey maps}

The total intensity maps show much stronger diffuse emission towards
the inner Galaxy ($l<60^{\circ}$) originating from spiral
arms\cite{srh+11b}. For the mixed emission in the inner region of the
Galactic plane, by using the multi-frequency total power data together
with our $\lambda$6\ cm map, we separated the thermal bremsstrahlung
emission component and the non-thermal synchrotron emission
component\cite{srh+11b}, we found that the thermal emission is about
60\% of the detected $\lambda$6\ cm emission. In the outer
Galaxy\cite{xhr+11}, the most outstanding region is the Cygnus complex
located at about $77^{\circ}<l<87^{\circ}$ and
$-2^{\circ}<b<4^{\circ}$, where indistinguishable radio emission from
{\sc Hii} regions and other objects accumulates in the tangential
direction of the local arm\cite{hhs09}. In the anti-centre region of
$l>120^{\circ}$, most prominent are large {\sc Hii} regions and
supernova remnants (SNRs) in the Perseus arm\cite{grh+10}.

When eliminating the ground radiation with polynomial fits as stated
above, polarized emission with spatial scale of the scan length of
each small survey area is filtered out. Absolute baselevels for $U$
and $Q$ are important to calculate correct polarization angles and
intensities. Missing large-scale components leads to a
misinterpretation of polarization data. For the $\lambda$6\ cm survey,
the ongoing C-Band All-Sky Survey (CBASS) might provide the missing
large-scale emission. However, CBASS data are not yet available. The
best we can do at present is to restore the large-scale $U$ and $Q$
components at $\lambda$6~cm using the WMAP K-band (22.8~GHz)
polarization data\cite{phk+07,hwh+09}. The spectral index of the
polarization intensity of the synchrotron emission in the plane varies
from $\beta=-3.1$ to $\beta=-2.7$. The $U$ and $Q$ maps of
high-latitude regions at 22.8~GHz were scaled to $\lambda$6~cm
(4.8~GHz) according to the spectral index $\beta$ to restore the
baselevel of $\lambda$6~cm $U$ and $Q$ maps, from which the $PI$ and
$PA$ maps are obtained.

From the polarization map of the survey area, we found many
polarization patches and depolarization structures of different
scales\cite{grh+10,srh+11b,xhr+11}. The prominent objects are again
large {\sc Hii} regions because of their depolarization and SNRs
because of their polarized emission. There are also many polarized
spurs and ``depolarized canals''. A few large polarized patches with
an angular size of a few degrees are probably discrete features within
1~kpc.  We have studied the fluctuation properties in the $U$, $Q$,
and $PI$ maps in the survey region of $10^{\circ}\leq l
\leq230^{\circ}$ by a structure function
analysis\cite{xhr+11}. Although the indices of the structure functions
are always close to about 0.5, the fluctuation power is much lower for
the anti-center region. More enhanced polarization structures on small
scales give much more fluctuation power towards the inner
Galaxy\cite{xhr+11}. The $\lambda$6\ cm polarization horizon in the
inner Galaxy is about 4~kpc\cite{srh+11b}, which means that the
observed polarized radiation comes from the Sagittarius and Scutum
arm.

\subsection{HII regions and Faraday screens}

Large {\sc Hii} regions are prominent radio sources in the
$\lambda$6\ cm total power map. In our polarization map, they often
appear due to their depolarization and Faraday rotation.

Along the periphery of many large {\sc Hii} regions (a few degrees in
size), very clear depolarization has been found in our restored
polarization map\cite{xhr+11}. This indicates either strong magnetic
fields or large electron density towards the boundaries of {\sc Hii}
regions. The regular magnetic fields in {\sc Hii} regions have been 
studied recently by using the rotation measure data\cite{hmg11}. 
A detailed modelling is in progress.

Polarization angles in some regions of degrees in size significantly
deviate from the surroundings\cite{shr+07,grh+10,xhr+11,srh+11b},
which are caused by foreground Faraday screens. These screens have a
very small electron density and therefore cannot be detected in total
intensity. Our modeling shows that the line of sight component of the
magnetic fields in the screens is up to tens of $\mu$G, much stronger
than the average interstellar magnetic fields. The nature of
Faraday screens remains unclear.
 
\subsection{Measurements of known supernova remnants}

Because of the intermediate beam size of the $\lambda$6\ cm
observation system, we have advantage to map large SNRs of several
degrees in size, which is a very ambitious task for large radio
telescopes with very small beams.

The first $\lambda$6\ cm test observation\cite{srh+06} was carried out
to the large SNR Cygnus loop (G74.0$-$8.5), with the apparent size of
$3^{\circ} \times 4^{\circ}$. We obtained a total intensity image,
from which we derived the integrated flux density at $\lambda$6\ cm
and the radio spectrum from 408~MHz to 4.8~GHz when combined with the
flux densities from the literature. The spectrum does not show any
steepening.  Our new $\lambda$6\ cm polarization image indicates a
distinct magnetic field configuration for the northern and southern
part of the SNR, which may support the idea for the Cygnus Loop 
resulting from two smaller SNRs\cite{ukb02}.

In the survey test region\cite{shr+07} of $122^{\circ}<l<129^{\circ}$,
we studied the SNRs G126.2+1.6 and G127.1+0.5. For the former we
excluded the suggested spectral break \cite{tl06} at about 1~GHz.
We have also studied 17 supernova remnants in the survey region which
are larger than $1^{\circ}$ in the fifth survey paper \cite{ghr+11a}
and 50 smaller ones in the seventh survey paper\cite{srr+11a}. Using the
Urumqi data, we determined the flux densities at $\lambda$6\ cm.
Together with the flux density data at other frequencies from
literature, we derived the radio spectra of these SNRs. A number of
the previously reported flux densities and spectral indices were
corrected/improved by our new measurements.

Some large SNRs are outside the survey region. We have also observed
them and studied individually. Besides the Cygnus Loop\cite{srh+06},
the faintest SNR in the Galaxy, G156.2+5.7\cite{xhs+07}, was imaged in
total intensity and polarization by Urumqi $\lambda$6\ cm and
Effelsberg $\lambda$11\ cm observations. A highly polarized central
patch is detected, and the toroidal magnetic field configuration is
identified for the SNR.
For the large SNR in the anti-center region of the Galactic plane,
S147, the high quality Urumqi $\lambda$6\ cm data confirm the spectral
break\cite{fr86} of the SNR at about 1.5~GHz\cite{xfrh08}. Two
populations of electrons with different energy spectral indices are
responsible for the diffuse emission and the filamentary emission
component.
Using the Urumqi $\lambda$6\ cm flux density data, together with
Effelsberg $\lambda$11\ cm and $\lambda$21\ cm data (see http: //www.mpifr.de /survey.html), we clarify that
SNR HB3 has a straight spectrum\cite{shg+08b} with a spectral index of $\alpha =
-0.61\pm0.06$ between 1.4~GHz and 4.8~GHz, which is consistent with
that at frequencies below 1~GHz\cite{kffu06} and rejected
the suggested spectral flattening above 1~GHz\cite{tl05}.
For another faint SNR G65.2+5.7, we obtained the integrated radio
spectral index from 83~MHz to 4.8~GHz\cite{xrfh09} and estimated the
magnetic field strength in the southern filamentary shell using 
new $\lambda$6\ cm and Effelsberg $\lambda$11\ cm polarization data.
Using the polarization maps of the Urumqi $\lambda$6\ cm and
Effelsberg $\lambda$11\ cm observations, together with data in
literature, the spectral index and its distribution as well as the
magnetic fields of the shell and central branch of SNR CTA~1 were
studied\cite{srw+11c}. A possible foreground Faraday screen partly
covering CTA~1 was identified from the rotation measure distribution.

In summary, of all ($\sim$ 80) known SNRs we investigated with the
Urumqi data, for 10\% of them polarization was newly detected and 35\%
of them the integrated flux densities were measured for the first time
at $\lambda$6~cm. The Urumqi data are at the highest frequency for
more than 50\% of the SNRs, which play an important role in
determining their spectral indices. The survey gives the first
(complete) polarization images for about 30\% of the SNRs, which is
important to understand the magnetic field structure of the SNRs.

Using the high-quality data of the Urumqi $\lambda$6\ cm and
Effelsberg $\lambda$11\ cm and $\lambda$21\ cm data, we have disproved
three objects, G192.8$-$1.1\cite{ghr+11a}, G166.2+2.5
(OA184)\cite{fks+06}, and G16.8$-$1.1\cite{srr+11a}, as being
SNRs. Their emission turns out to be thermal without any
polarization, and hence they are very probably {\sc Hii} regions.

\subsection{Discoveries of new SNRs and HII regions}

Sensitive surveys always lead to new discoveries, so does the Urumqi
$\lambda$6\ cm survey.  With the high sensitivity of the
$\lambda$6\ cm system, we were able to discover two new shell-type
SNRs: G178.2$-$4.2 and G25.1$-$2.3\cite{gsh+11b}.  G178.2$-$4.2 has a
size of $72'\times 62'$ with strongly polarized emission along its
northern shell. The spectral index of G178.2$-$4.2 is $\alpha =
-0.48\pm0.13$.  Its surface brightness is $\Sigma_{\rm 1~GHz} =
7.2\times10^{-23}{\,\rm Wm^{-2} Hz^{-1} sr^{-1}}$, which ranks it the
second faintest known Galactic SNR. G25.1$-$2.3 is revealed by its
strong southern shell, which has a size of $80'\times 30'$. It has a
non-thermal radio spectrum with a spectral index of $\alpha
=-0.49\pm0.13$.

We have also identified a number of {\sc Hii} regions.  The first is
G124.9+0.1 ($30'$) in the test survey region\cite{shr+07}, which acts
as a Faraday screen. Its regular magnetic field parallel to the
line-of-sight can thus be estimated to be about 3.9~$\mu$G. Other new
{\sc Hii} regions are G12.8$−$3.6 ($100'\times24'$)\cite{srh+11b},
G56.7$−$0.6 ($65'\times27'$)\cite{srh+11b}, G98.3$-$1.6
($90'\times66'$)\cite{xhr+11}, G119.6+0.4
($44'\times44'$)\cite{xhr+11}, G148.8+2.3
($63'\times40'$)\cite{ssh+08a}, G149.5+0.0 ($30'$)\cite{ssh+08a} and
G169.9+2.0 ($27'\times20'$)\cite{ssh+08a}. A more detailed study of
{\sc Hii} regions from the survey is being prepared (Gao et al. in
prep.)

\section{Conclusions} 

Our new polarization survey at $\lambda$6\ cm can delineate the
Galactic magnetic fields in both large and small scales, in a larger
and deeper area than that observed at low radio frequencies. With its
high quality data, we studied the radio properties of many known SNRs
covered by the survey. We discovered two new SNRs with very faint
surface brightness. We revealed several prominent Faraday screens with
high RM values, {\sc Hii} regions, and many interesting
polarization/depolarization structures, e.g. depolarization around
the edge of ionized {\sc Hii} regions, canals, and polarization
patches.

\section*{Acknowledgments}
{\small We thank Drs. Hui Shi and Chen Wang for conducting
  some observations. The Sino-German $\lambda$6\ cm polarization
  survey was carried out with a receiver system constructed by
  Mr. Otmar Lochner at MPIfR with financial support by the MPG and the
  NAOC. We thank Dr. Peter M\"uller for the installation and
  adaptation of Effelsberg data reduction software at the Urumqi
  observatory, and Mr. Maozheng Chen and Jun Ma for operation support
  and maintenance, and also Prof. Ernst F\"urst for a lot of help and
  involvement in the project. The survey team was supported by the
  National Natural Science foundation of China (10773016, 10821061,
  and 10833003).}

\end{document}